\documentclass[twocolumn]{aastex62}
\usepackage{lineno}
\usepackage{threeparttable}
\usepackage{amsmath}
\usepackage{subfigure}
\usepackage[utf8]{inputenc}
\DeclareUnicodeCharacter{0301}{\dash}
\DeclareRobustCommand\dash{%
  \unskip\nobreak\thinspace\textemdash\allowbreak\thinspace\ignorespaces}

\newcommand{\um}{$\mu$m}

\newcommand{\kms}{km\,s$^{-1}$}
\newcommand\lya{Ly$\alpha$}

\newcommand{\oiii}{[O\thinspace{\sc iii}]}

\newcommand{\ci}{[C\thinspace{\sc i}]}

\newcommand{\civ}{[C\thinspace{\sc iv}]}
\newcommand{\heii}{[He\thinspace{\sc ii}]}
\newcommand{\mkms}{\rm {km\,s^{-1}}}

\newcommand\coft{$\rm CO(4-3)$\,}
\newcommand\cott{$\rm CO(3-2)$\,}

\newcommand\cooz{$\rm CO(1-0)$\,}
\newcommand\Sa{$Source\,A$}
\newcommand\Sb{$Source\,B$}
\newcommand\Sc{$Source\,C$}
\graphicspath{{./}{figures/}}

\accepted{September 21, 2021}

\shorttitle{Sample article}
\shortauthors{Li et al.}

\begin{document}
\title{\MakeTextUppercase{Discovery of a protocluster core associated with an enormous Ly$\alpha$ Nebula at $z = 2.3$}}

\correspondingauthor{Ran Wang}
\email{rwangkiaa@pku.edu.cn}

\author[0000-0002-3119-9003]{Qiong Li}
\affil{Department of Astronomy, School of Physics, Peking University, Beijing 100871, P. R. China}
\affil{Kavli Institute for Astronomy and Astrophysics, Peking University, Beijing, 100871, P. R. China}
\affil{Department of Astronomy, University of Michigan, 311 West Hall, 1085 S. University Ave, Ann Arbor, MI, 48109-1107, U.S.A.}
\author[0000-0003-4956-5742]{Ran Wang}
\affil{Kavli Institute for Astronomy and Astrophysics, Peking University, Beijing, 100871, P. R. China}
\author[0000-0001-7147-3575]{Helmut Dannerbauer}
\affil{Instituto de Astrofisica de Canarias (IAC), E-38205 La Laguna, Tenerife, Spain}
\affil{Universidad de La Laguna, Dpto. Astrofísica, E-38206 La Laguna, Tenerife, Spain}
\author[0000-0001-8467-6478]{Zheng Cai}
\affil{Department of Astronomy, Tsinghua University, Beijing 100084, China}
\author[0000-0003-2983-815X]{Bjorn Emonts}
\affil{National Radio Astronomy Observatory, 520 Edgemont Road, Charlottesville, VA 22903, USA}
\author[0000-0002-7738-6875]{Jason Xavier Prochaska}
\affil{Department of Astronomy and Astrophysics, University of California, 1156 High Street, Santa Cruz, California 95064, USA}
\affil{University of California Observatories, Lick Observatory, 1156 High Street, Santa Cruz, California 95064, USA}
\author[0000-0002-4770-6137]{Fabrizio Arrigoni Battaia}
\affil{Max-Planck-Institut fur Astrophysik, Karl-Schwarzschild-Str 1, D-85748 Garching bei M\"unchen, Germany}
\author[0000-0002-7176-4046]{Roberto Neri}
\affil{Institut de Radioastronomie Millimétrique (IRAM), 300 rue de la piscine, F-38406 Saint-Martin-d'Hères, France}
\author[0000-0001-6469-1582]{Chengpeng Zhang}
\affil{Department of Astronomy, School of Physics, Peking University, Beijing 100871, P. R. China}
\affil{Kavli Institute for Astronomy and Astrophysics, Peking University, Beijing, 100871, P. R. China}
\author[0000-0003-3310-0131]{Xiaohui Fan}
\affil{University of Arizona (Steward Observatory), USA}
\author{Shuowen Jin}
\affil{Instituto de Astrofisica de Canarias (IAC), E-38205 La Laguna, Tenerife, Spain}
\affil{Universidad de La Laguna, Dpto. Astrofísica, E-38206 La Laguna, Tenerife, Spain}
\author{Ilsang Yoon}
\affil{National Radio Astronomy Observatory, 520 Edgemont Road, Charlottesville, VA 22903, USA}
\author{Shane Bechtel}
\affil{Department of Physics, University of California, Santa Barbara, CA 93106, USA}

\begin{abstract}
The MAMMOTH-1 nebula at $z=2.317$ is an enormous \lya\, nebula (ELAN) extending to a $\sim$\,440 kpc scale at the center of the extreme galaxy overdensity BOSS 1441. In this paper, we present observations of the \cott\, and 250 GHz dust-continuum emission from the MAMMOTH-1 using the IRAM NOrthern Extended Millimeter Array. Our observations show that \cott\, emission in this ELAN has not extended widespread emission into the circum- and inter-galactic media. We also find a remarkable concentration of six massive galaxies in \cott\, emission in the central $\sim$100 kpc region of the ELAN. Their velocity dispersions suggest a total halo mass of $M_{200c} \sim 10^{13.1} M_{\odot}$, marking a possible protocluster core associated with the ELAN. The peak position of the \cott\, line emission from the obscured AGN is consistent with the location of the intensity peak of MAMMOTH-1 in the rest-frame UV band. Its luminosity line ratio between the \cott\, and \cooz\, $r_{3,1}$ is 0.61$\pm$0.17. The other five galaxies have \cott\, luminosities in the range of (2.1-7.1)$\times 10^9$ K $\mkms$ pc$^2$, with the star-formation rates derived from the 250GHz continuum of ($<$36)-224 $M_{\odot}$yr$^{-1}$. Follow-up spectroscopic observations will further confirm more member galaxies and improve the accuracy of the halo mass estimation.
\end{abstract}



\section{Introduction}\label{section1}
Enormous \lya\, nebulae (ELANe)
are rare and bright (SB$_{\rm Ly\alpha}>$10$^{-17}$\,erg\,s$^{-1}$\,cm$^{-2}$\, arcsec$^{-2}$) Ly$\alpha$-emitting regions extending up to hundreds of kpcs (e.g., \citealt{Cantalupo2014,Cai2017b,Fab2018,Cai18, Cai19}).
They host multiple active galactic nuclei \citep{Hennawi2015,Fab2018a}, and reside in overdense environments as seen from the Lyman-alpha emitters (LAEs) around them \citep{Hennawi2015,Cai2017}.
A major question is how the star-formation or AGN activities are fueled and evolve within the ELANe. To answer this question, cold molecular gas was mapped in the fields of these ELANe \citep[e.g.][]{yang2012,yang2014,Wagg2012,Ao2020,Decarli2021}.

The ELAN around the Spiderweb Galaxy ($\sim$250 kpc) at $z$=2.2 \citep{Miley2006} revealed the first evidence for the existence of a cold molecular circumgalactic medium (CGM) of the ELANe in the distant universe by using the Australia Telescope Compact Array (ATCA) \citep{Emonts2016}. \cooz\ observations show a massive ($\sim$70 kpc, $M_{\rm H2}\sim10^{11} M_{\odot}$) reservoir of gas in the CGM that cooled well beyond the temperature of Ly$\alpha$-emitting gas ($T\sim 10^4$\,K), and is actively feeding star formation across the halo.
\coft\, and \ci\, are detected across $\sim$50 kpc, following the distribution of previously detected low-surface-brightness \cooz\, across the CGM \citep{Emonts2018}.
Its line ratio and carbon abundance are similar to that of the Milky Way and star-forming galaxies (SFGs) \citep{Emonts2018}.
Thus, observations of the CO emission from the ELANe has the potential (i) to probe the physical conditions and kinematics of the cold gas component within the system and (ii) to constrain the energy and momentum output released by the star-formation and/or AGN activities into the interstellar/circumgalactic medium (ISM/CGM). Furthermore, such observations will address the nature of the central source and the mechanism that powers these gaseous nebulae.

The ELAN MAMMOTH-1 is an enormous \lya\, nebula discovered by \citet{Cai2017b}, extending on $442$ kpc scale. It is also the first radio-quiet source to have strongly extended ($\sim$ 30 kpc) \civ\ and \heii\ emission \citep{Cai2017b}.
MAMMOTH-1 resides in an extremely overdense field, BOSS1441, containing strong Ly$\alpha$ absorptions at $z=2.32\pm0.02$ in the spectra of five background QSOs, projected within 20 $h^{-1}$ Mpc scale \citep{Cai2017}.
\citet{Cai2017} confirmed that the LAE overdensity ({$\delta_{\rm{LAE}}=\rho_{\rm{LAE}}/<\rho_{\rm{LAE}}>-1$}) in the  MAMMOTH-1 field is $10.8\pm2.0$ on 15 Mpc, which could be one of the most overdense fields found to date.
\citet{Fab2018} also revealed an overdensity of submillimeter galaxies (SMGs) of $\delta_{\rm SMG} = 3.0$ around the peak area of this LAE overdensity by using the Submillimetre Common-User Bolometer Array-2 (SCUBA2) on the James Clerk Maxwell Telescope.
Its oberved frame 350GHz (850$\mu$m) continuum detection suggests the far-infrared (FIR) luminosity of $L_{\rm FIR}=2.4\times10^{12} L_{\odot}$ from ELAN MAMMOTH-I by assuming a dust temperature of 45\,K and an emissivity index of 1.6 \citep{Fab2018}.
The optical, IR, and submillimeter observations of the ELAN MAMMOTH-1 suggest that this ELAN could be powered by an extreme system with massive star formation and strong AGN activity in the middle of a massive large-scale structure.
Furthermore, \citet{Emonts2019} detected \cooz\, luminosity of  $L'_{\rm CO(1-0)}\sim3.8\pm0.8\times10^{10}$\,K\,\kms\,pc$^2$ from the ELAN MAMMOTH-I, revealing a molecular gas mass of  $M_{\rm H2}\sim 1.4(\alpha_{\rm CO}/3.6)\times10^{11}\,{M_{\odot}}$. Strikingly, 50\% of the \cooz\, spans $\sim30$ kpc into the CGM.

In this work, we present IRAM NOrthern Extended Millimeter Array (NOEMA) observations of the \cott\, line and 250\,GHz continuum emission. These provide us further information about the dense gas and star forming activity in and around this ELAN.
We describe the observations and data reduction in \S\ref{section2}, and present the results in \S\ref{section3}.
Then we report the newly discovered galaxy group (a protocluster core) within this system and discuss the properties of the molecular gas from the galaxy members in \S\ref{section4}. We conclude with a brief summary in \S\ref{section5}.
Throughout this paper, we assume a flat cosmological model with
$\Omega_{\Lambda} = 0.7, \Omega_{m} = 0.3$, and $H_0 = 70\, \mkms \, \rm Mpc^{-1}$.
Finally, we note that based on the flux peak of the Ly$\alpha$ emission, \citet{Cai2017} defined the center of the MAMMOTH-I Nebula as \Sb. Follow-up studies of MAMMOTH-I (e.g., \citealt{Emonts2019} and \citealt{Fab2018}) adopt its terminology.
Here, we follow this naming convention; see also Table~\ref{table1} and \ref{table2}.

\section{NOEMA observations}\label{section2}
\subsection{\cott\, transition}
The ELAN MAMMOTH-1 was observed with NOEMA (ID: S18CW) centered on ($\alpha_{2000}$,
$\delta_{2000}$) = ($14^{h}41^{m}24.47^{s}$, $+40^\circ$03\arcmin09.67\arcsec) between 2018 November to 2019 January in C configuration with 10 antennas.
The total observing time is 8 hours and the on-source time is 4.6 hr.
We use the 3~mm receiver and the correlator PolyFix in dual polarization mode, tuning one of the 3.9 GHz basebands on the observed frequency 104.250 GHz of the redshifted \cott transition.

We use the quasars 1505+428, 1504+377 and J1438+371 as phase and amplitude calibrators.
The RF calibrators were 2013+370 and 3C273, the flux calibrator was MWC349.
In the 3~mm band (band 1 of NOEMA), the absolute flux calibration is accurate within 10\%. The calibrated visibility data were imaged with the software package MAPPING (part of GILDAS), using natural baseline weighting and the Hogbom cleaning algorithm.
The final synthesized beam sizes are $2\farcs3 \times 1\farcs6$.
To further check if there are missing CO intensities in more extended area, we tapered and re-weighted the visibilities to a lower angular resolution of $3\farcs2 \times 3\farcs2$.
The field of view (primary beam) for NOEMA observations at $\nu_{obs}$ = 100GHz are 50\arcsec.
Then we made the primary beam correction.
The maximum recoverable scale at the observed frame 100GHz is roughly 41\arcsec.
The 1$\sigma$ rms sensitivity of the natural-weighted image cube is 0.3 mJy beam$^{-1}$ per 45 $\mkms$\, channel, while the rms of the tapered-image cube is 0.4 mJy beam$^{-1}$ per 45 $\mkms$\, channel.
We mainly focus on the small beam size ($2\farcs3 \times 1\farcs6$) in the following discussion.

\subsection{250 GHz dust continuum}

We imaged the observed frame 250 GHz continuum emission of the MAMMOTH-1 field with NOEMA in C configuration (ID: W19CX). The observations were carried out on 2019 December 29 and 31, using the PolyFiX correlator with the full available continuum bandwidth of 15.5 GHz in dual polarization.  The flux scale was calibrated on MWC349 and the phase was checked with the calibrator 1505+428 close to our target. The total observing time is 8 hr with 5 hr on-source. The primary beam for NOEMA observations at $\nu_{\rm obs}$= 250 GHz ($\nu_{\rm rest}\sim$ 828.2 GHz) are 21\arcsec. The FWHM synthesized beam size is $0\farcs87 \times 0\farcs65$ and the position angle is 49$^\circ$. The final continuum 1$\sigma$ rms sensitivity in the central region of the cleaned image is 0.04 mJy\,beam$^{-1}$.  Primary beam correction was applied to the final flux measurements described in Section~\ref{section3.4}.

\subsection{Ancillary data}
The rest-frame optical deep image of the MAMMOTH-I field was obtained using the Hubble Space Telescope ($HST$) Wide Field Camera 3 (WFC3) with the F160W filter (ID: 14760). The observation was carried out in Cycle 24 with an exposure time of $\sim$2665 sec. we use nine HST/WFC3 pointings to cover this central region in nine orbits to measure the detailed rest-frame optical morphology.
The data reduction is conducted using Multidrizzle \citep{Koekemoer2003}, and the detailed procedures follow the descriptions in \citet{Cai2016}. To optimize the output data quality, we choose a final output pixel scale of 0\farcs06 instead of the initial pixel scale 0\farcs13 and final pixfac parameter 0.7 (shrinking pixel area) after different trials of combinations of parameters.

We also observed the Ly$\alpha$ line of MAMMOTH-I with the Keck Cosmic Web Imager (KCWI) on the Keck II telescope of the W.\,M.~Keck Observatory in Hawaii in 2018 May (seeing $\sim$ 1\farcs5). The on-source exposure time is 1 hour. We used the Blue Medium Grating with the Large Slicer (slice width $\sim3\farcs5$), resulting in a spectral resolution of 2000 and field of view of $33\arcsec \times20\arcsec$, centered on \Sb\,, the quasar optical position given by \citet{Cai2017}.
To convert the spectral images and calibration frames (arcs, flats, bias) to a calibrated data cube, we used the IDL-based KCWI data reduction pipeline\footnote{Available at \url{https://github.com/kcwidev/kderp/releases/tag/v0.6.0}}. Basic CCD reduction is performed on each science frame to obtain a bias-subtracted, cosmic-ray-cleaned and gain-corrected image. The continuum flat images are employed for CCD response corrections and pixel-to-pixel variations. We used a continuum-bar image and an arc image (ThAr) to define the geometric transformations and wavelength calibration, generating a rectified object data cube (see the pipeline documents\footnote {\url{https://github.com/kcwidev/kderp/blob/master/AAAREADME}}). Twilight flats were used for slice-to-slice flux correction, and the data were corrected for atmospheric refraction. Each object and sky frame was flux calibrated with the standard star BD28d4211.
For each exposure we found the QSO centroid to measure the offsets between exposures and then performed a weighted mean with inverse-square variance weighting to construct the final data cube.

The \cooz\, line data of this field was observed using the Very Large Array (VLA) in the most compact D-configuration and published in \citet{Emonts2019}, which is included in the analysis of this paper.
The exposure time is 14 hr on-source, with an rms noise level of 0.057 mJy beam$^{-1}$ channel$^{-1}$, for a channel width of 30 $\mkms$.
The field of view (primary beam) for VLA observations at $\nu_{\rm obs}\sim$ 34.81 GHz ($\nu_{\rm rest}$ = 115.27 GHz) are 1.3\arcmin.
The synthesized beam is 2\farcs$6\times$2\farcs3.
We use these near-infrared and radio data to compare with our \cott observations in order to study the molecular gas distribution and kinematics and thus provide a link between the ELAN and the stellar build-up of this system.

\section{Results}\label{section3}
\begin{figure*}
\centering
\includegraphics[width = \linewidth]{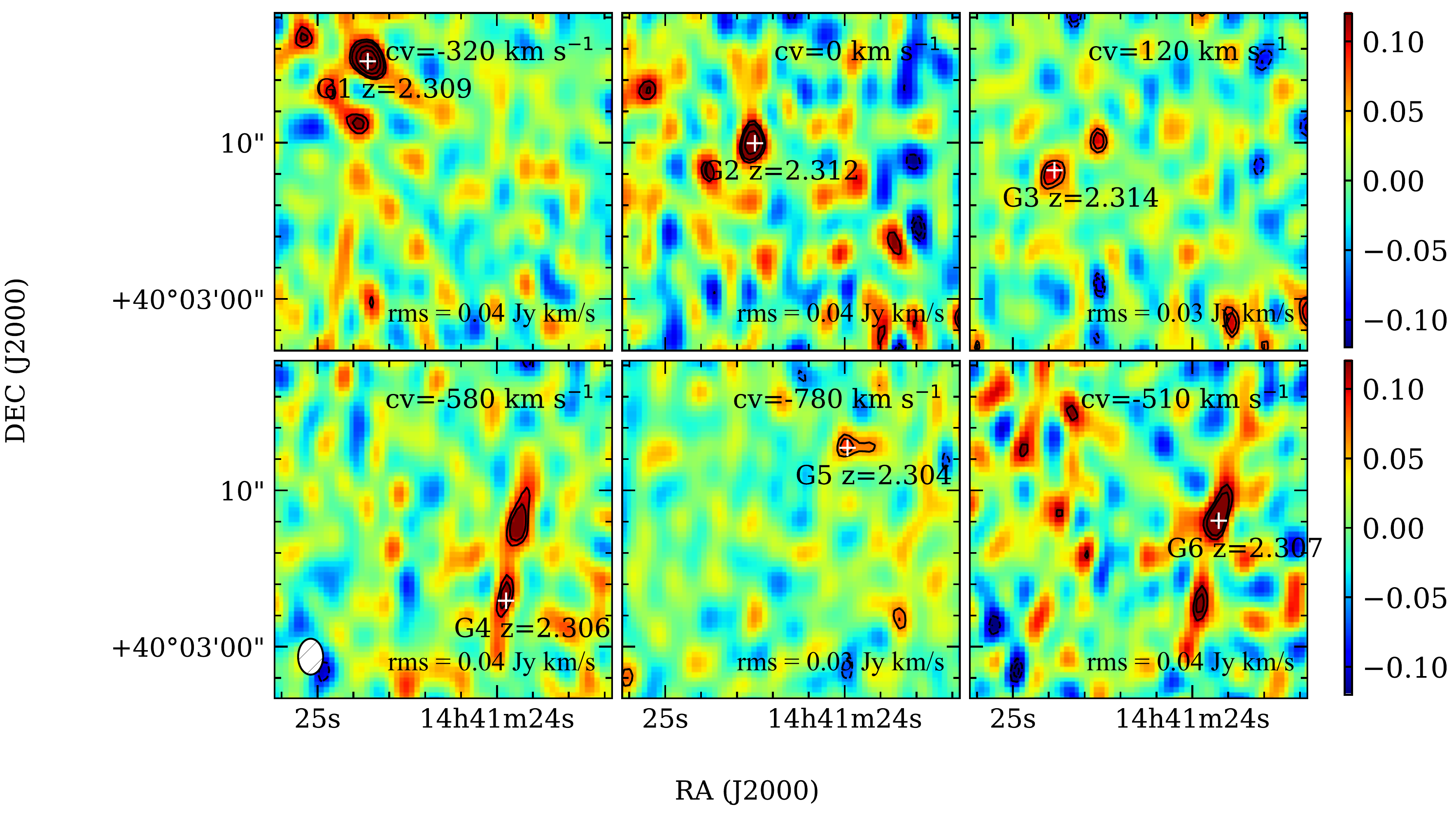}
\caption{The velocity-integrated \cott\, intensity map ($\sim$20\arcsec $\times$ 20\arcsec) of 3$\sigma$ detected sources around the ELAN MAMMOTH-I in NOEMA observations. The contour levels are [-3, -2.5, +2.5, +3, +4, +5...]$\times \sigma$, with $\sigma$ = [0.03-0.04] Jy beam$^{-1}$ $\mkms$. The map has been made the primary beam correction. The integrated velocity ranges are shown in the yellow channels on the right side of Fig. \ref{figure1}. The center velocity (`cv') is the offset velocity of the \cott\, peak emission respect to G2 (source B). The rms is also marked on the individual map.
}
\label{figure1_new}
\end{figure*}

\begin{figure*}
\centering
\includegraphics[width = \linewidth]{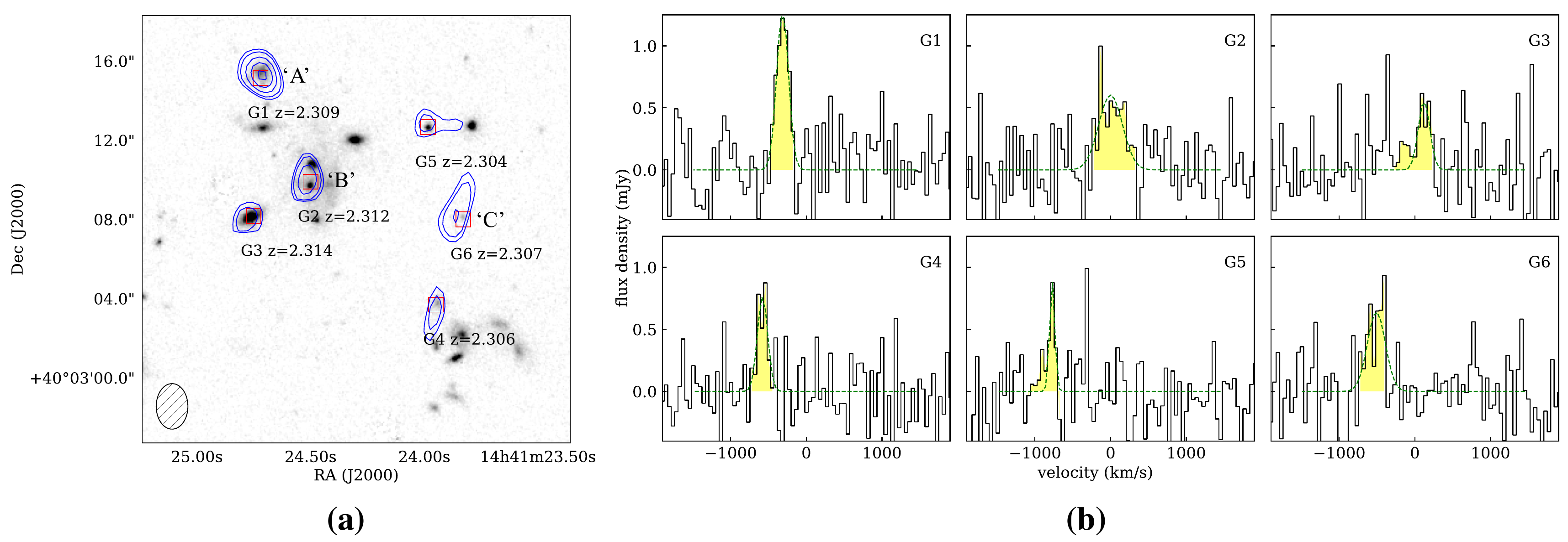}
\caption{
\cott\, emission of the galaxies detected around the ELAN MAMMOTH-I (approximately centered on G2)
using NOEMA.
{\bf(a)} The left panels show the members of the galaxy group ($\sim$20\arcsec $\times$ 20\arcsec) confirmed by \cott\, emission, overlaid on the HST F160W image. The \cott\, intensity map has been made the primary beam correction. The contour levels are [+2.5, +3, +4, +5...]$\times \sigma$, with $\sigma$ = [0.03-0.04] Jy beam$^{-1}$ $\mkms$. The NOEMA beam ($2\farcs3 \times 1\farcs6$) is denoted by the ellipse.
{\bf(b)} The right panels show the one-dimensional spectra of these CO emission-line detections, extracted from the detection peak (marked as the square in the left panel). The root-mean-square (rms) noise level is 0.3 mJy beam$^{-1}$ per 45 $\mkms$ channel. \Sb\/ (G2) is the center of the Ly$\alpha$, the zero velocity of all the spectra corresponds to Source `B' at $z = 2.3123$. The area filled with colors indicates the regions where emission is detected. The green dashed lines are the gaussian fit.
}
\label{figure1}
\end{figure*}

Within the central 16\arcsec\ region of ELAN MAMMOTH-1, we detect strong \cott\, emission in six proximate galaxies. The velocity-integrated CO intensity maps are shown in Fig.~\ref{figure1_new}.
These sources are all $>3\sigma$ detections in \cott\, and have HST optical counterparts.
Three of them are also detected in \cooz\, with the VLA \citep{Emonts2019}.
Fig.~\ref{figure1} shows the intensity map of the \cott\, emission and the spectra extracted from the peak of the \cott\, detections.
The intensity map is the integrated intensity over the line-emitting region. The selected velocity ranges for each source are shown in the yellow channels in the right side of Fig.~\ref{figure1}.
In the right panels of Fig.~\ref{figure1}, we did not detect any continuum; and we fit the spectra with a Gaussian profile, then calculated the line center, full width half maximum (FWHM) and line flux.
The derived results are summarized in Table~\ref{table1}. The spectroscopic redshifts of the sources ($z_{\rm spec} = 2.3037 - 2.3137$, in a range from $-770$ to $+120$ \kms\,) are all consistent with being physically related to the same structure at $z = 2.3$.

\subsection{\cott\, line detections of individual sources}

At the center of the ELAN MAMMOTH-1, the \cott\, line flux of $Source\,B$ (G2) is $0.237\pm0.051$ Jy \kms (Fig.~\ref{figure1}).
We fit the FWHM$_{\rm CO(3-2)}$ to be 370$\pm90$\,\kms, which is a typical CO line width found in samples of SFGs and quasars (e.g., \citealt{Carilli2013,Ueda2014}).
However, the line width of the \cooz\, detected in \citet{Emonts2019} is much narrower, only 85 \kms\,.
We re-analyze the previous \cooz\, data, fitting it with a double Gaussian profile with no constraint applied. After subtracting the narrow Gaussian component, the line width of the new broad \cooz\, component is similar to that of \cott\,. More details are presented in Section~\ref{section4.2}.
The redshift derived from \cott\, line is $z_{\rm CO} = 2.3123\pm0.0006$, while the redshift derived from \lya\, is $z_{Ly\alpha}=2.329\pm0.013$ \citep{Cai2017}.
This shift may be due to resonant scattering of Ly$\alpha$.
The 2D Gaussian fit for the velocity-integrated image suggests line emission from an unresolved source.

Fig.~\ref{figure1} also reveals two bright sources in \cott\, in the nearby region of $Source\,B$, labeled as $Sources\,A$ and $C$. They were detected in \cooz\, line emission down to a sensitivity level of 0.057 mJy beam$^{-1}$ channel$^{-1}$ for a channel width of 30 $\mkms$ \citep{Emonts2019}.
Adopting the naming convention from \citet{Emonts2019}, $Source\,A$ (G1) is the brightest detection in both \cott\, and \cooz\, lines in these systems.
However, it is not located inside the ELAN MAMMOTH-I (6\arcsec away from $Source\,B$) and has its own \lya\, emission (Li in prep.).
It is also unresolved. The line flux I$_{\rm CO(3-2)}$ is $0.298\pm0.044$\,Jy \kms.
Its FWHM$_{\rm CO(3-2)}$ of $\sim 180$\kms\, is similar to that of \cooz ($\sim 170$\kms).
$Source\,C$ (G6) is a faint detection roughly 9\arcsec\/ west of $Source\,B$  with a line flux of $0.245\pm0.056$Jy \kms.
The line width is FWHM$_{\rm CO(3-2)}=280\pm70$\,\kms, similar to the \cooz\, of $\sim230$\,\kms.
The velocity offset relative to $Source\,B$ is $\sim$510 \kms.

The other three sources (G3, G4, G5) are all $>3\sigma$ detections in \cott\, and have HST optical counterparts. However, they are not detected in \cooz\,. By applying the line ratio of $r_{3,1}$ = 0.52 for SFGs \citep{Kirkpatrick2019}, the 1$\sigma$ rms of the VLA CO luminosity measurement of  L$^{\prime}_{\rm CO(1-0)}\sim0.3\times10^{10}$ K km s$^{-1}$ pc$^2$ would correspond to L$^{\prime}_{\rm CO(3-2)}\sim1.6\times10^{9}$ K km s$^{-1}$ pc$^2$ for \cott\,.
The sensitivity of the VLA observations is insufficient to detect the \cooz\, line from these three objects.

\subsection{The distribution of the \cott emission}\label{section3.2}
\begin{figure*}
\centering
\includegraphics[width = \linewidth]{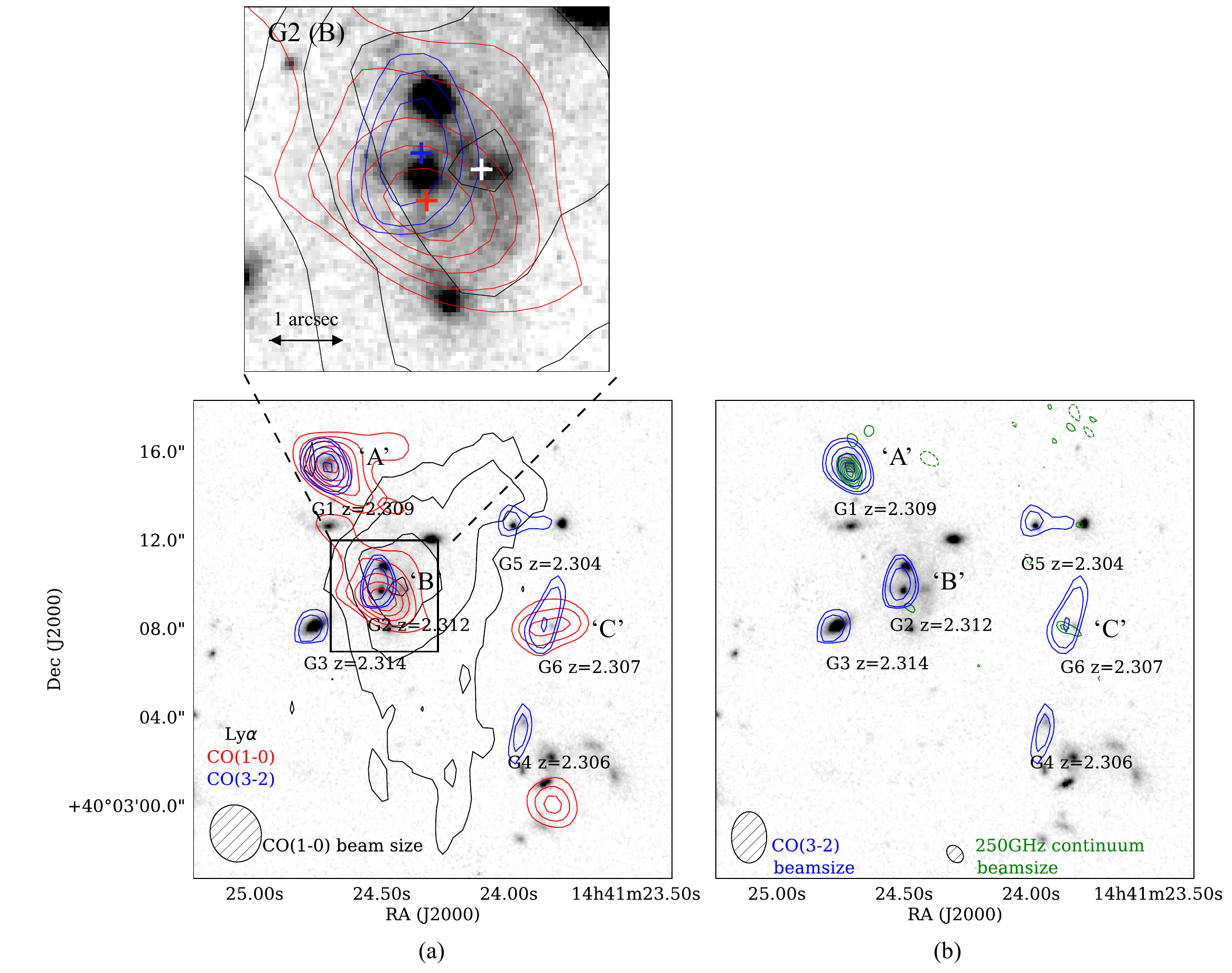}
\caption{{\bf(a)} \cott\, emission from NOEMA (blue contours), \cooz\, from VLA (red contours) and Ly$\alpha$ from KCWI (black contours, Li et al. in prep.) overlaid onto the HST/WFC3 FW160W image (Cai et al. in prep.).
Six sources of this galaxy group (or a so-called protocluster core) are detected with NOEMA; three of them are also detected in \cooz\, with the VLA.
The Ly$\alpha$ contours levels are [+2.5, +7.5, +22.5, +45.0]$\times$ 0.04 mJy\,beam$^{-1}$. The \cooz\, and \cott\, contours levels are [+2.5, +3.5,...]$\times \sigma$.
The synthesized beam of \cooz\, is 2\farcs$6\times$2\farcs3, the synthesized beam of \cott\, is 2\farcs$3\times$1\farcs6. The synthesized beam at 250GHz is 0\farcs$87\times$0\farcs75.
The upper panel shows a zoom around \Sb\,. The white, blue and red crosses indicate the peak of \lya\,, \cott\,, and \cooz\,.
{\bf(b)} 250~GHz dust continuum and \cott\, emission from NOEMA (green and blue contours) overlaid onto the HST/WFPC3 image. Three galaxies in the galaxy group have above 3 sigma continuum detections. The 250 GHz continuum contours levels are [-4,-3,+3,+4,+5,+6,...]$\times \sigma$. The continuum sensitivity is 0.04 mJy\,beam$^{-1}$.
}
\label{figure2}
\end{figure*}

Now we compare CO, \lya\,, and optical counterparts of this system. In Fig.~\ref{figure2} we show the map of the \cott\, line emission from NOEMA (blue contours), \cooz\, from VLA (red contours) and the Ly$\alpha$ from KCWI (black contours, Li et al. in prep) overlaid on the HST/WFC3 FW160W image (Cai et al in prep). The HST image shows four optical counterparts around $Source\,B$, see Fig. \ref{figure2}(a). The \cott\, and \cooz\, peaks of $Source\,B$ (G2) are almost consistent with the central HST optical source. The peak of \lya\, is located at another optical counterpart. The CO peak and \lya\, peak have an offset of $\sim1.5$\arcsec ($\sim$13kpc).
Taking into account the accuracy of the radio interferometric positions\footnote{The radio interferometric position measurement uncertainty of \Sb\, ($\Delta\theta\sim0.5\frac{\theta_{beam}}{SNR}$, \citealt{Reid1988}) using NOEMA and VLA are $\sim$0\farcs2 and 0\farcs2, respectively} from the VLA and NOEMA, and the seeing of the Ly$\alpha$ observations of $\sim1\farcs0$, the offset could be real.

$Source\,A$ (G1) has an optical counterpart in the HST image.
It was also detected \lya\, emission, 7\arcsec\, away from the \lya\, peak of the ELAN MAMMOTH-I. The \cott peak of $Source\,A$ (G1) coincides with \cooz. \citet{Emonts2019} reported that the \cooz emission of $Source\,A$ and $Source\,B$ have extended features of $\sim$25 kpc and 30 kpc, respectively. However, our \cott\, observations suggest that they are unresolved,
which implies that \cott is compact and emitted from the star-forming region within the galaxy.

$Source\,C$ (G6) does not have any \lya\, counterpart but shows an HST optical detection.
$Source\,C$ also has \cooz\, emission in VLA observations \citep{Emonts2019}.
The \cooz\, and \cott\, lines also show that it is a point source. Its \cott\, line flux is weaker than for A and B.

The maximum recoverable scale is $\sim41$\arcsec\, for our \cott\, observations, corresponding to the physical scale of $\sim$340 kpc at $z\sim2.3$.
Here we tapered the beam to recover the additional emission in CGM with a beam size of 3\farcs2$\times$3\farcs2, corresponding to the physical scale of $\sim$27 kpc.
We assumed the same average $r_{3,1}$=0.6 and the same CO line width in the CGM, the sensitivity of \cooz\, in VLA data is 0.057 mJy beam$^{-1}$ per 30 $\mkms$\, channel and the derived sensitivity of \cott\, is 0.5 mJy beam$^{-1}$ per 45 $\mkms$\, channel. Our NOEMA data sensitive is 0.4 mJy beam$^{-1}$ per 45 $\mkms$\, channel, enough to detected the \cooz\, extended features in \cott\,, but at this sensitivity in our NOEMA data, we did not detect diffuse \cott\, emission.
This indicates that there is no diffuse \cott\, line emission across the nebula. It is different from the diffuse \cooz emission which appears extended across a region of $\sim$30 kpc.

\subsection{Line ratios}
To constrain the nature of the sources of this system, here we calculate the \cott\, line luminosity of each source as
${L}_{\rm CO}^{'}=3.25\times {10}^{7}\times{I}_{\rm CO} {D}_{\rm L}^2 (1+z)^{-3} {\nu}_{\rm obs}^{-2}\ \rm K\ km\ s^{-1}\ pc^{2}$ \citep{Solomon1992} where ${I}_{\rm CO}$ is the integrated line flux in Jy \kms, ${D}_{\rm L}$ is the luminosity distance in Mpc and ${\nu}_{\rm obs}$ is the observing frame \cott\, line frequency. The derived line luminosities are in the range of (2.1$-$7.1)$\times 10^9\, \rm K\ km\ s^{-1}\ pc^{2}$.
The \cott/\cooz luminosity ratios ($r_{3,1}$) of $Sources\,A$, $B$ and $C$ are 0.59$\pm$0.17, 0.61$\pm$0.17 and 0.54$\pm$0.25, respectively.
The median line ratios $r_{3,1}$ for AGN- and star-formation dominated galaxies are 0.92$\pm$0.44 and 0.52$\pm$0.17 \citep{Kirkpatrick2019}.
\citet{Carilli2013} suggested the average line ratio $r_{3,1}$ of quasars is 0.97. $Source\,B$ is lower than this value and lies toward the low end of the expected AGN-dominated range.

Low $r_{3,1}$ in ELANe have also been reported previously. e.g., \citet{Genzel2003} observed SMM\,J02399, a BAL quasar in a $>140$\,pkpc ELAN at $z\sim2.8$ \citep{Ivison1998,LiQ2019}. Its $r_{3,1}$ is $0.48\pm0.13$.
By contrast, the central radio galaxy MRC 1138-262 in the ELAN Spiderweb Galaxy (\lya\, extended $\sim200$~kpc, \citealt{Miley2006}) shows a very high global $r_{4,1}$($L'_{\rm CO(4-3)}/L'_{\rm CO(1-0)})$ of $1.00\pm0.28$ \citep{Emonts2018}. Its CGM has $r_{4,1}=0.45\pm0.17$, similar to SFGs.

Regardless of the presence of AGNs, the ratios $r_{3,1}$ in different galaxies vary greatly from 0.4 to 0.9.
At high $z$, the literatures, such as \citet{Harris2010}($z=2.5-2.9$ SMGs) and \citet{Aravena2010}($z\sim1.5$ normal star-forming galaxies), report $r_{3,1}\sim0.6$.
\citet{Sharon2016} reports that there is no statistically significant difference in the mean line ratio ($r_{3,1} = 0.90\pm0.40$ for both populations combined) in $z\sim2$ galaxies including both AGNs and SMGs.
Furthermore, \citet{Riechers2020} report $r_{3,1}=0.84\pm0.26$ in $z=2-3$ main-sequence galaxies from the ASPECS surveys.
The higher-J CO transition observations, at least $J_{\rm up}\geq4$, can further reveal the excited gas and especially excitation from AGNs.

\subsection{250 GHz continuum and FIR luminosity}\label{section3.4}

Previous continuum observation of MAMMOTH-1 using SCUBA-2 at the observed frame 350GHz revealed bright dust continuum emission with a flux of $S_{\rm 350GHz}=4.6\pm0.9$ mJy \citep{Fab2018}. However, the SCUBA-2 beam size of 15\arcsec is too large to constrain the continuum emission from individual galaxies. All the CO-detected objects in this area could contribute to the SCUBA-2 flux. Three continuum sources are detected at $>4\sigma$ in this field from our NOEMA 250GHz (1.2mm at observed frame) map at $0\farcs87 \times 0\farcs65$ resolution. They are counterparts of the CO detections described above (Sources A, B, and C). We used 2D Gaussian fitting to measure the continuum and detect a continuum flux of 0.74$\pm$0.15 mJy from Source A. The continuum emission is marginally resolved along the major axis with a deconvolved source size of $(1.25\pm0.32)\arcsec\times(0.57\pm0.26)\arcsec$. This is the brightest continuum detection in the MAMMOTH-1 field.  We detect a 4$\sigma$ continuum source at the position of Source B. The continuum emission is unresolved and we adopt the peak surface brightness as the total continuum flux which is 0.14$\pm$0.03 mJy.  We also detect continuum emission from Source C with a flux of 0.18$\pm$0.05 mJy. The source is unresolved as well.
The continuum detections are summarized in Table~\ref{table1}.

\begin{figure*}
\centering
\includegraphics[width = \linewidth]{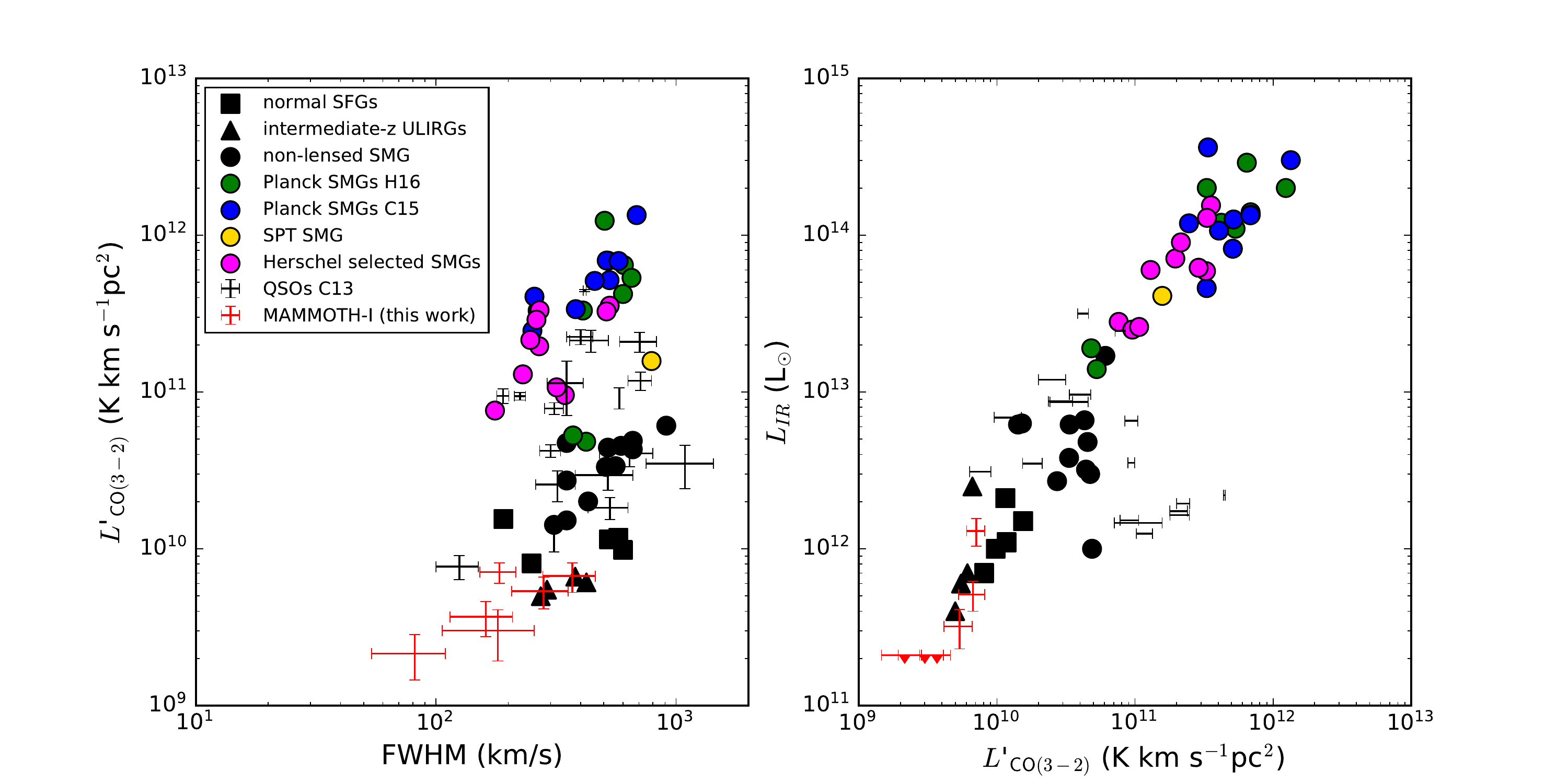}
\caption{{\bf Left:} The relation between FWHM and $L'_{\rm CO(3-2)}$ of the high-redshift galaxies and quasars. The galaxies in MAMMOTH-I are shown in red. The comparable quasar sample is from \citet{Carilli2013}. The comparable galaxies from the literature are as follows: intermediate-z ULIRGs (black filled triangles: \citealt{Magdis2014}), normal SFGs (black filled squares: \citealt{Daddi2015,Arabsalmani2018}), nonlensed SMGs (black filled circles: \citealt{Bothwell2013}), Herschel-selected lensed SMGs (magenta filled circles: \citealt{Riechers2013,YangOmont2017}), SPT-selected lensed SMG (gold filled circle; \citealt{Strandet2017}), Planck-selected SMGs from \citep{Harrington2016}(green filled circles) and \citealt{Canameras2015}(blue filled circles). {\bf Right:} the relation between $L_{\rm IR}$ and $L'_{CO(3-2)}$.
}
\label{figure3}
\end{figure*}

The NOEMA continuum map at 250GHz allows us to determine the FIR luminosity and the star-formation rate. We assume a modified blackbody for optically thin thermal dust emission, with the dust temperature of 42K for \Sb\,(AGN), and 35~K for other galaxies.
We adopt an emissivity index of $\beta=1.6$
\citep{Beelen2006}.
For the non-detections, we adopt a 3$\sigma$ upper limit.
We estimate FIR luminosity by integrating the modified blackbody function in 8-1000 \um\ anchored by 250GHz flux measurement.
The derived FIR luminosities of \Sa\,, \Sb\, and \Sc\, are (13.0$\pm$2.6), (5.1$\pm$1.1) and (3.2$\pm$0.9)$\times 10^{11}L_{\odot}$. In Fig.~\ref{figure3}, we compared these galaxies around ELAN MAMMOTH-I with other quasar samples, intermediate-$z$ ULIRGs, normal SFGs and SMGs \citep{Carilli2013,Magdis2014,Daddi2015,Arabsalmani2018,Bothwell2013,Riechers2013,YangOmont2017,Strandet2017,Harrington2016,Canameras2015,Dannerbauer2019}.
The FIR luminosities of the galaxies in MAMMOTH-I are slightly lower than that of quasars but still consistent with normal SFGs, see Fig.~\ref{figure3} right panel).
The CO-FIR luminosity ratios for the MAMMOTH-I members are at the lower end but still follow the trend of the relation between $L_{\rm IR}$ and $L'_{\rm CO(3-2)}$.
This indicates that these galaxy members are all following the star- formation law, in which the star-formation rate traced by the IR luminosity has a tight relation with the mass of fuel traced by \cott\,.

The derived total FIR luminosity of the whole nebula from the 250~GHz observations is $<2.7\times10^{12} L_{\odot}$.
\citet{Fab2018} reported a bright detection at observed frame 350 GHz (850$\mu$m), with a flux of $S_{\rm 350 GHz}=4.6\pm0.9$ mJy (with a beam size of 15\arcsec) and a 3$\sigma$ upper limit of about 16 mJy at observed frame 667GHz (450$\mu$m).
The group of six galaxies are all covered by the 350GHz\, SCUBA-2 beam ($\sim$15\arcsec). The total flux at the observed frame 250GHz is $<$1.42\,mJy.
Assuming the modified blackbody for \Sb\, (AGN; $T_{\rm dust}=42$K, $\beta=1.6$) and  other galaxies ($T_{\rm dust}=35$K, $\beta=1.6$), the derived total 350GHz\, flux should be $<3.7$\,mJy. The results are slightly lower than the SCUBA2 detections. This may be due to the standard matched filter applied in the SCUBA2 data reduction in order to increase the point source detectability \citep{Fab2018}.

Assuming a star-formation-powered emission, we estimate the star formation rate as SFR/$M_{\odot}$\,yr$^{-1}$ $= 4.5\times 10^{-44}\times L_{\rm FIR}$/erg\,s$^{-1}$ \citep{Kennicutt1998}; see Table~\ref{table2}.
The three sources detected at 250~GHz are violent starbursts with a star-formation rate ranging between 54 - 224 M$_{\odot}$\,yr$^{-1}$.

\section{Discussion}\label{section4}
\subsection{Halo mass of this protocluster core}
MAMMOTH-I resides in the density peak of the large-scale structure BOSS1441, which is one of the most overdense fields discovered to date with an LAE density of 12 times higher than that of a random field density on a 15\,cMpc scale.
We discovered a remarkable galaxy concentration at a redshift of $z\sim2.3$, containing six gas-rich galaxies spectroscopically confirmed through the \cott\, transition in the central $\sim$100 kpc region.
The high density of massive galaxies and velocity dispersion of this overdensity suggest that it could be embedded in a collapsed, cluster/group-sized halo.
In this section, we further explore the total halo mass of this galaxy group.

Galaxy cluster velocity dispersion provides a reliable estimate of the cluster mass \citep{Evrard2008,Munari2013,Saro2013,wangtao2016}.
The cluster redshift is $z=2.308$, determined by the weighted average of the \cott\, redshifts of these six galaxies.
The galaxy proper velocities $v_i$ are then derived from their redshifts $z_i$ by $v_i = c(z_i - z)/(1 + z)$ \citep{Danese1980}.
The line-of-sight velocity dispersion $\sigma_v$ is the square root of the weighted sample variance of proper velocities \citep{Beers1990,Ruel2014} which is estimated to be $\sigma_v = 320 \, \mkms$.
We assume that only the inner portion of this protocluster is virialized. Using the relation between velocity dispersion and total mass suggested in \citet{Evrard2008},
$\sigma_{\rm DM}(M,z) = \sigma_{\rm DM,15}(h(z)M_{200c}/10^{15}M_{\odot})^\alpha$,
we derive a total halo mass of ELAN MAMMOTH-I of $M_{200c} \sim 10^{13.1} M_{\odot}$ (by using the canonical value of $\sigma_{\rm DM,15}\sim 1083 \mkms$ and the logarithmic slope $\alpha \sim 0.33617$).
It is an upper limit if the system has not yet virialized.
The previous studies of AGNs, SMGs and bright LABs show that they are expected to live in halos of 10$^{12-14}$ $M_{\odot}$ (e.g., \citealt{White2012,wangtao2016,wilkinson2017,YangYujin2010}).
The total halo mass of the ELAN MAMMOTH-I is in a good agreement with that of AGNs and SMGs.

Previous work on protocluster cores based on sensitive observations of CO and ionized carbon \citep{Miller2018,Oteo2018,Gomez-Guijarro2019} reported a total molecular gas masses of $\sim 10^{11}M_{\odot}$ and the total halo mass as high as $\sim 10^{13}M_{\odot}$. We use $L'_{\rm CO(1-0)}$ from \citet{Emonts2019} to derive the cold molecular gas mass $M_{\rm H2}$ for \Sa\,, \Sb\, and \Sc\,. The gas masses are in the range of (3.6-4.3)$\times10^{10}M_{\odot}$. Here we assume a typical conversion factor for high-$z$ galaxies of $\alpha_{\rm CO}$ = $M_{\rm H2}$/$L'_{co}$ = 3.6 $M_{\odot}$ (K km s$^{-1}$ pc$^2$)$^{-1}$ (e.g., \citealt{Daddi2010,Genzel2010}). For the other three galaxies without detections of \cooz\, we assume a gas excitation similar to the typical SFGs with $r_{3,1}$ = 0.52. Their gas masses derived from $L'_{\rm CO(3-2)}$ in our NOEMA observations are in the range of (1.5 - 2.6)$\times10^{10}M_{\odot}$, which is listed in Table \ref{table2}. The total molecular gas mass in the MAMMOTH-1 protocluster is 1.8$\pm0.1\times10^{11}M_{\odot}$. In summary, the gas, dust, and stellar properties of MAMMOTH-I are all comparable to these protocluster cores mentioned before, indicating that MAMMOTH-I should be the progenitor of a galaxy cluster.

Finally, we note that due to the influence of large-scale structure in and around clusters \citep{WhiteMartin2010}, there are uncertainties in the estimation of the mass for an individual cluster based on velocity dispersion.
We are also aware that the sample used to estimate the velocity dispersion only includes SFGs.
The quiescent galaxies would also change the estimate of the velocity dispersion \citep{wangtao2016}.
As we cannot rule out completely the existence of quiescent galaxies of this galaxy cluster, we are planning to confirm more member galaxies spectroscopically to further improve the accuracy of the velocity dispersion estimation.

\subsection{The CGM within MAMMOTH-1}\label{section4.2}
Extended \cooz emission has been found on the scale of tens of kpc around high-$z$ massive galaxies or in protoclusters (e.g., \citealt{Emonts2014,Emonts2016,Dannerbauer2017}). For ELAN MAMMOTH-I,
\citet{Emonts2019} also reported that half of the cold molecular gas traced via the \cooz transitions stretches on $\sim$30 kpc into the CGM, appearing to be a wide tail of gas in $Source\,A (G1)$ and an extended reservoir of cold gas in $Source\,B (G2)$.
Extended high-$J$ CO emission in the CGM also has been reported in some metal-line nebulae. \citet{Ginolfi2017} detected a large structure of molecular gas reservoir traced by \coft\, in an \oiii\, nebula, extended over 40 kpc.
Strikingly, with our NOEMA observations, we did not find any evidence of
extending \cott emission.

This could be due to the fact that the critical density of CO$(J,J- 1)$ scales roughly with ${n_{\rm crit}} \propto J^3$. The ground-transition \cooz\, has an effective critical density of only several 100 cm$^{-3}$, and the $J$ = 1 level of CO is substantially populated down to $T\sim10$ K.
However, these values increase by an order of magnitude or more for the high-$J$ transitions, like \cott\, and higher.
Compared to \cooz\,, the \cott\, is more likely a tracer of the star forming region or compact gas around an AGN. As in such regions, the gas is warmer and denser compared to that in the CGM. CO molecules tend to populate at high-J levels, and the optical depth also increases.

In the CGM around $Sources\,A$ and $B$, the molecular gas is not excited to give strong \cott\, emission.
The ELANe Slug ($z = 2.282$) and Jackpot ($z = 2.041$) also do not show any extended \cott\, emission \citep{Decarli2021};
whereas, for the ELAN around the Spiderweb Galaxy, \coft\, (and \ci\, ) are detected across $\sim$50 kpc, comprising $\sim30\%$ of the total flux.
The Spiderweb Galaxy has a massive, cold molecular gas reservoir in the CGM that is roughly twice as luminous as that seen in \cooz\, in MAMMOTH-I \citep{Emonts2016, Emonts2019}.
In addition, the Spiderweb Galaxy emits jets of relativistic particles visible in radio observations and has a metal-enriched outflow \citep{Pentericci1997,Nesvadba2006}. By contrast, MAMMOTH-1 does not have a radio jet and does not show any clear features of outflow in \civ and \heii\, observations (Zhang et al. in prep). These could explain the reason why \cott\, around ELAN MAMMOTH-I is only associated with the galaxies but not with the CGM.

\begin{figure*}
\centering
\includegraphics[width = \linewidth]{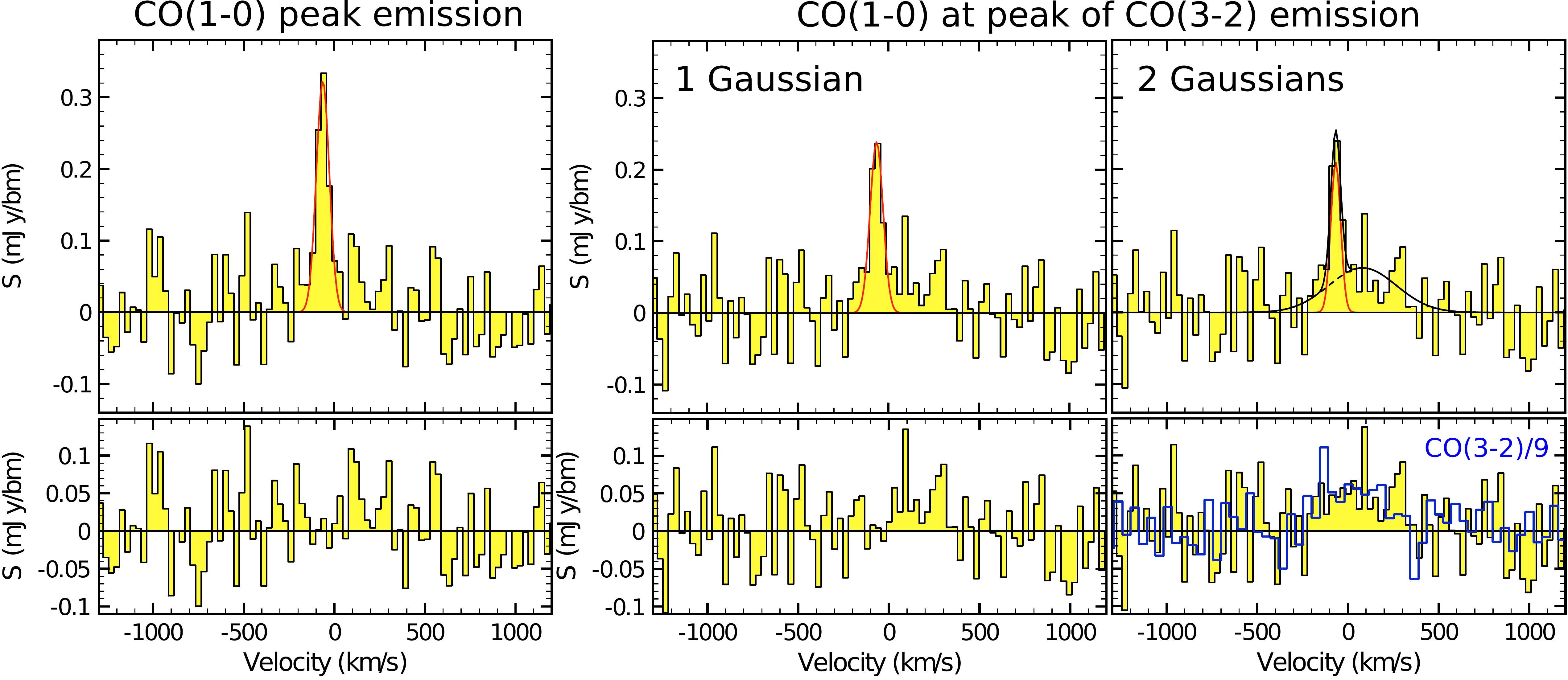}
\caption{The intensity spectra of \cooz\, in region B, taken at the location of the peak of the \cooz\,(left), and at the location of the peak of the \cott\, emission (right). These locations are marked with the red and blue `+' sign in Fig. \ref{figure2}, respectively. The spectra are not mutually independent, because the both locations fall within the same synthesized beam in the \cooz\, data, meaning that features with sufficient signal-to-noise appear in both spectra. The spectrum on the left is fitted with a single Gaussian function from \citet{Emonts2019}. On the right, the top panels show the same spectrum, with the first panel showing a single Gaussian with width constrained to that of the \cooz\, peak emission (FWHM\,=\,85 km\,s$^{-1}$), and the second panel showing a double Gaussian fit with no constraints applied. The bottom panels show the residuals after subtracting the narrow (red) Gaussian component from each spectrum. In the bottom right-hand panel, a weak, broad \cooz\, component is present at the peak of the \cott\, emission. Overlaid in blue is the \cott\, spectrum, scaled down by a factor of nine for easy comparison.}
\label{figure5}
\end{figure*}

In Fig.~\ref{figure5} we revisit the \cooz\, analysis of $Source\,B$ from \citet{Emonts2019} and compare its \cott\, and \cooz\, transitions.
The \cott\, line shows a broad line profile with FWHM of $\sim350$\,\kms.
It is comparable to the typical line width of SMGs (e.g., \citealt{Carilli2013,Bothwell2013,Goto2015}), which implies that \cott\, traces the molecular gas from the star forming regions within the galaxy.
By comparison, while the \cooz signal in $Source\,B$ is dominated by a narrow \cooz line (FWHM of 85\,\kms) outside the central galaxies, at the location of the peak of the \cott emission there appears to be an additional weak, broad component. This broad \cooz component is detected only at the 3.2$\sigma$ level when minimizing the contribution of the narrow component (rightmost panel of Fig. \ref{figure5}). However, its properties are remarkably similar to the \cott spectrum in $Source\,B$ (G2), with FWHM\,=\,440\,$\pm$\,160 km\,s$^{-1}$, $z$\,=\,2.3132\,$\pm$\,0.0009, $I_{\rm CO(1-0)}$\,=\,0.029\,$\pm$0.015 Jy km s$^{-1}$, and $r_{3,1}$\,=\,0.9\,$\pm$\,0.5. The line ratio value is closer to AGN-dominated molecular gas emission. Therefore, the VLA data are consistent with the presence of two \cooz features in $Source\,B$ (G2), namely a CGM component of dynamically cold gas with very low velocity dispersion and excitation conditions (previously described in \citealt{Emonts2019}), as well as a weak counterpart to the \cott emission of molecular gas associated with the central galaxies. However, for the latter, Fig. \ref{figure5} shows that ambiguities in line-fitting remain and hence deeper \cooz observations are needed before drawing any firm conclusions.

\section{Summary}\label{section5}
In this paper, we present IRAM NOEMA observations of the \cott\, line and dust continuum at 250\,GHz of the ELAN MAMMOTH-1. 
We discovered a remarkable galaxy concentration, containing 6 massive galaxies in the central $\sim$100 kpc region, forming a so-called protocluster core.
The total halo mass derived from the velocity dispersion is $M_{200c} \sim 10^{13.1} M_{\odot}$. For this ELAN, we did not detect not any extended widespread \cott\, emission on CGM scales down to our sensitivity of 0.4 mJy beam$^{-1}$ per 45 $\mkms$\, channel. Our finding suggests that the \cott, emission traces the warmer and denser molecular gas, heated through star formation. Future spectroscopic follow-up observations will confirm more member galaxies in order to better constrain the nature of the ELAN MAMMOTH-1 and its environment which should evolve most likely into a galaxy cluster.

\acknowledgments
{\bf Acknowledgement:}
This work is based on observations carried out under project numbers S18CW and W19CX with the IRAM NOEMA Interferometer. IRAM is supported by INSU/CNRS (France), MPG (Germany), and IGN (Spain). This work made use of GILDAS\,\footnote{GILDAS: http://www.iram.fr/IRAMFR/GILDAS}, a collection of state-of-the-art software oriented toward (sub-)millimeter radio-astronomical applications (either single-dish or interferometer).
R. W. acknowledges support from the Thousand Youth Talents Program of China and the National Science Foundation of China (NSFC) grants No. 11473004 and 11721303.
H. D. acknowledges financial support from the Spanish Ministry of Science, Innovation, and Universities (MICIU) under the 2014 Ramón y Cajal program RYC-2014-15686 and under the AYA2017-84061-P, co-financed by FEDER (European Regional, Development Funds), and in addition, from the Agencia Estatal de Investigación del Ministerio de Ciencia e Innovación (AEI-MCINN) under grant (La evolución de los cúmulos de galaxias desde el amanecer hasta el mediodía cósmico) with reference (PID2019-105776GB-I00/DOI:10.13039/501100011033).

\begin{table*}
\caption{IRAM NOEMA measurements of the ELAN MAMMOTH-I.}\label{table1}
\centering
\begin{threeparttable}
\vspace{0.2cm}
\begin{tabular}{lcccccccc}
\hline \hline \noalign {\smallskip}
 Source     &RA       &DEC   &redshift  &\rm $\Delta v_{\rm CO}$  &FWHM$_{\rm CO}$   &$I_{\rm CO(3-2)}$    & S$_{\rm 250~GHz}$    \\
            &J2000    &J2000 &          &km s$^{-1}$      &km s$^{-1}$   &Jy km s$^{-1}$ & mJy \\
   (1)      &    (2)  &  (3) & (4)      & (5)             &   (6)        &    (7)  & (8)\\
\hline \noalign {\smallskip}
G1(A) &14:41:24.72&+40:03:15.14 &2.3088 $\pm$ 0.0004 &-310 &180$\pm$30 &0.298$\pm$0.044 &0.74$\pm$0.15$^{*}$ \\
G2(B) &14:41:24.50&+40:03:09.90 &2.3123 $\pm$ 0.0006 &0    &370$\pm$90 &0.237$\pm$0.051 &0.14$\pm$0.03$^{**}$  \\
G3    &14:41:24.75&+40:03:08.17 &2.3137 $\pm$ 0.0004 &120  &180$\pm$80 &0.113$\pm$0.040 &$<0.12$ \\
G4    &14:41:23.95&+40:03:03.69 &2.3059 $\pm$ 0.0004 &-580 &160$\pm$50 &0.182$\pm$0.046 &$<0.12$   \\
G5    &14:41:23.98&+40:03:12.66 &2.3037 $\pm$ 0.0003 &-770 & 80$\pm$30 &0.092$\pm$0.030 &$<0.12$  \\
G6(C) &14:41:23.83&+40:03:08.00 &2.3067 $\pm$ 0.0005 &-500 &280$\pm$70 &0.245$\pm$0.056 &0.18$\pm$0.05 \\
\hline \noalign {\smallskip}
\end{tabular}
\small \textbf{Notes.} \\
Col~(1): Source name. Cols.~(2) and (3):  R.A. and Declination in J2000. Col.~(4): spectroscopic redshift derived from the \cott\, observations. Col~(5): the offset velocity of the \cott\, peak emission, respect to source B. Col~(6): the FWHM of \cott\,, derived by a Gaussian fitting to the \cott\, profile. Col~(7): \cott\, integrated velocity intensity. The intensity and luminosity errors are derived from the fitting of the emission-line profile of the CO peak. Col~(8): 250~GHz dust-continuum observations, upper limits for G3, G4 and G5 are 3$\sigma$.\\
* G1 is marginally resolved at the 250GHz continuum map along the major axis (1.25$\pm$0.32)\arcsec\,$\times$(0.57$\pm$0.26)\arcsec\,. The continuum source position is 14:41:24.71 +40.03.15.13.\\
** A $>$4$\sigma$ peak close to G2 (central AGN of MAMMOTH-I): the 250GHz continuum peak position is 14:41:24.48 40:03:08.98, about 0.7\arcsec\, away from the phase center.
\end{threeparttable}
\end{table*}

\begin{table*}
\caption{Physical properties of the protocluster core MAMMOTH-I.}\label{table2}
\centering
\begin{threeparttable}
\vspace{0.2cm}
\begin{tabular}{lcccccccccc}
\hline \hline \noalign {\smallskip}
 Source &L$_{\rm FIR}$&SFR            &$L'_{\rm CO(1-0)}$       &$L'_{\rm CO(3-2)}$ & $r_{3,1}$   &M$_{\rm gas}$\\
        & 10$^{11}$ $L_{\odot}$ &$M_{\odot}$\,yr$^{-1}$ &10$^{10}$ K km s$^{-1}$ pc$^2$& 10$^9$ K km s$^{-1}$ pc$^2$  &  &10$^{10}$ $M_{\odot}$\\
   (1)  &    (2)  &  (3) & (4)      & (5)             &   (6)        &    (7)    \\
\hline \noalign {\smallskip}
G1(A) &13.0$\pm$2.6 &224$\pm$45   &1.2$\pm$0.3  & 7.1$\pm$1.1  &0.59$\pm$0.17    &4.3$\pm$0.1 \\
G2(B)-narrow &5.1$\pm$1.1  &88$\pm$19    &1.1$\pm$0.2  & 6.7$\pm$1.4  &0.61$\pm$0.17    &4.0$\pm$0.1 \\
G2(B)-broad  &  &    &0.7$\pm$0.4  &   &0.91$\pm$0.51    & \\
G3    &$<$2.1       &$<$36        &$<$0.9       &3.0$\pm$1.1   &$>$0.33          &2.1$\pm$0.8 \\
G4    &$<$2.1       &$<$36        &$<$0.9       &3.7$\pm$0.9   &$>$0.41          &2.6$\pm$0.6 \\
G5    &$<$2.1       &$<$36        &$<$0.9       &2.1$\pm$0.6   &$>$0.24          &1.5$\pm$0.4 \\
G6(C) &3.2$\pm$0.9  &54$\pm$15    &1.0$\pm$0.4  &5.4$\pm$1.2   &0.54$\pm$0.25    &3.6$\pm$0.1 \\

\hline \noalign {\smallskip}
\end{tabular}
\small \textbf{Notes.} \\
Col.~(1): Source name. Col.~(2): FIR luminosity from 8 to 1000 \um, assuming a modified blackbody for optically thin thermal dust emission, with a dust temperature of 42K for \Sb\, (as it is a Type-II AGN) and 35K for other galaxies. We adopt an emissivity index of $\beta=1.6$, which is the typical value found in FIR-bright quasars at $z \sim$ 2$-$4 \citep{Beelen2006}. Given upper limits are 3$\sigma$. Col.~(3): star-formation rate. Here we use SFR = 4.5$\times 10^{-44}\times L_{\rm FIR}$ \citep{Kennicutt1998}. Col.~(4): \cooz\, luminosity of the ELAN MAMMOTH-I from \citet{Emonts2019}. Col.~(5): \cott\, luminosity from our NOEMA observations. Col~(6): \cott/\cooz line ratio. The fitted broad and narrow components of G2 are discussed in Section~\ref{section4.2}, which trace the center galaxy and the extended CGM gas,  respectively. Col~(7): molecular gas mass, assuming a typical conversion factor for high-z galaxies of $\alpha_{\rm CO}$ = $M_{\rm H2}$/$L'_{co}$ = 3.6 $M_{\odot}$ (K km s$^{-1}$ pc$^2$)$^{-1}$ (e.g., \citealt{Daddi2010,Genzel2010}). Gas masses for G1, G2 and G6 are derived from $L'_{\rm CO(1-0)}$ \citep{Emonts2019}. We assume that the other three galaxies are star forming-dominated galaxies $r_{3,1}$ = 0.52 and derive gas mass from $L'_{\rm CO(3-2)}$.
\end{threeparttable}
\end{table*}

\clearpage
\bibliography{ref} 



\end{document}